\documentclass[11pt,a4paper]{article}
\usepackage{jcappub}

\def\cm{\textrm{cm}}
\def\km{\textrm{km}}
\def\sec{\textrm{s}}

\def\Mpc{\textrm{Mpc}}
\def\eV{\textrm{eV}}
\def\MeV{\textrm{MeV}}
\def\GeV{\textrm{GeV}}
\def\TeV{\textrm{TeV}}
\def\PeV{\textrm{PeV}}
\def\yr{\textrm{yr}}
\def\Msun{\textrm{M}_{\odot}}
\def\kms{\km\ \sec^{-1}}
\def\ga{\gtrsim}
\def\la{\lesssim}
\def\nodata{...}
\def\ergps{\textrm{ergs}~\textrm{s}^{-1}}

\title{Olber's Paradox for Superluminal Neutrinos: Constraining Extreme Neutrino Speeds at TeV--ZeV Energies with the Diffuse Neutrino Background}
\author[a,b]{Brian C. Lacki}
\affiliation[a]{Jansky Fellow}
\affiliation[b]{Institute for Advanced Study,\\Einstein Drive, Princeton, New Jersey, USA}
\emailAdd{brianlacki@ias.edu}

\abstract{The only invariant speed in special relativity is $c$; therefore, if some neutrinos travel at even tiny speeds above $c$, normal special relativity is incomplete and any superluminal speed may be possible.  I derive a limit on superluminal neutrino speeds $v \gg c$ at high energies by noting that such speeds would increase the size of the neutrino horizon.  The increased volume of the Universe visible leads to a brighter astrophysical neutrino background.  The nondetection of ``guaranteed'' neutrino backgrounds from star-forming galaxies and ultrahigh energy cosmic rays (UHECRs) constrains $v/c$ at TeV--ZeV energies.  I find that $v/c \la 820$ at 60 TeV from the nondetection of neutrinos from star-forming galaxies.  The nondetection of neutrinos from UHECRs constrains $v/c$ to be less than 2500 at 0.1 EeV in a pessimistic model and less than 4.6 at 4 EeV in an optimistic model.  The UHECR neutrino background nondetection is strongly inconsistent with a naive quadratic extrapolation of the OPERA results to EeV energies.  The limits apply subject to some caveats, particularly that the expected pionic neutrino backgrounds exist and that neutrinos travel faster than $c$ when they pass the detector.  They could be improved substantially as the expected neutrino backgrounds are better understood and with new experimental neutrino background limits.  I also point out that extremely subluminal speeds would result in a much smaller neutrino background intensity than expected.}


\begin{document}
\maketitle\flushbottom

\section{Introduction}
An early question with implications for cosmology is Olber's Paradox: Why is the sky dark at night?  In a static, infinite Universe every line of sight should eventually intercept the surface of a star, and so the sky should be as bright as a stellar surface (see \cite{OlbersHistory} for a critical review of the early history of Olber's paradox).  The resolution of the paradox arises from the finite age of the Universe in Big Bang cosmology: stars only radiate for finite times, limiting the energy density of the background; in terms of flux, the Universe has a finite age and so we see only those galaxies where the light has had time to reach us \cite{OlbersSolution}.\footnote{In addition, while the Universe once was as bright as a stellar surface, the cosmic expansion redshifted that radiation to microwave frequencies (e.g., \cite{Peacock}).  However, the cosmic expansion plays a relatively minor role in setting the energy density or energy flux of \emph{starlight} \cite{OlbersSolution}.}  The background flux from the sky therefore depends on the speed $v$ of the particle, which sets the length scale of the ``horizon'' for that particle.

The recent claim of superluminal neutrinos ($\Delta v / c \equiv v/c - 1  \approx 2 \times 10^{-5}$ at $\sim 17\ \GeV$) from the OPERA experiment \cite{Adam11} has excited much interest.  Superluminal particles have been predicted to arise through Lorentz invariance violation or propagation through extra dimensions.  Several constraints on superluminal particles exist.  The most powerful limits use the kinematics of superluminal particles to infer new energy loss processes which would limit the range of particles, or alter known particle reactions.  Such limits have previously been derived for protons \cite{Coleman97} and photons \cite{Stecker01}.  Recent limits of this kind for neutrinos include those from $e^+e^-$ pair production \cite{NuToePair,NuToePairPRL,KinematicLimits,ICARUS}, neutrino splitting \cite{NuSplitting}, and suppression of pion decay \cite{KinematicLimits,NuVsPionDecay}.  These limits rule out the OPERA results in standard Lorentz invariance theories, and they become stronger at high energy.  But it is conceivable that these limits can be evaded, for example, if photons and electrons have the same speed $v \gg c$ at high energies as neutrinos \cite{NuToePairPRL}, if neutrinos transform into a superluminal particle that does not couple with the Standard Model sector \cite{DarkFTLNu}, or in some proposed modifications of Lorentz invariance that alter the laws of energy-momentum conservation \cite{FTLinDSR} or the spacetime metric \cite{Bimetric}. 

Generic limits arise from timing considerations if particles are detected from an astrophysical transient \cite{AstroLIVTests}.  In particular, gamma-ray bursts last only a few seconds in our frame and are billions of light years away; even a small difference in propagation speed will accumulate into huge differences in arrival times.  \emph{Fermi}-LAT observations of gamma-ray bursts has already severely constrained velocities of the form $\Delta v / c \propto E$ for photons from the detection of GeV $\gamma$-rays coincident with $\gamma$-ray bursts \cite{FermiLIVLimit}. 

At TeV energies, the Universe becomes opaque to $\gamma$-rays, but neutrinos can also provide powerful constraints if they are ever detected \cite{GRBNuLIVLimit,DecayOfAstroNus}.  Unfortunately, no astrophysical neutrinos have been detected except from the supernova SN1987A and the Sun.  The detection of core collapse neutrinos from SN1987A a few hours before it became visible in light provides strong constraints on $v/c$ at tens of MeV \cite{SN1987ALIVLimit}.

In special relativity, any speed minimally faster than light in one frame can appear much faster than light in another frame, suggesting that once we allow the possibility of superluminal particles at all, we should consider particles much faster than $c$.  Indeed, some ideas for superluminal neutrinos predict that the neutrino speed should grow at higher energy.  Higher order effects may cause $\Delta v / c$ to asymptote to some small limit, but once we allow the possibility of superluminal particles, there may be nothing special about the photon speed $c$ \emph{per se} so that high energy neutrinos travel at speeds much faster than $c$.  For all we know, higher order terms may cause $\Delta v / c$ to grow even faster at some special energy scale (e.g., exponentially) and asymptote to $v \gg c$.  This suggests we should look at higher energies where superluminal speeds could be huge.

Furthermore, if neutrino speed limits from SN1987A and the results of the OPERA experiment apply to all flavors, then $\Delta v / c$ must grow quickly at least between the MeV scales of SN 1987A's neutrinos and the GeV scales of the OPERA neutrinos \cite{SN1987AvsOPERA}.  Assuming $\Delta v / c \propto E^2$, the OPERA result implies $\Delta v / c \approx 1$ at 3 TeV; if $\Delta v /c \propto E$, then the OPERA result implies $\Delta v / c \approx 1$ at 0.7 PeV.  But regardless of whether the OPERA experiment result is correct, any $\Delta v/c$ that can be directly measured at low energies may be extrapolated to $v \gg c$ at high energies with a strong enough energy dependence.  

I present a constraint on extreme superluminal neutrino speeds at very high energies, under certain assumptions described in the next section.  In short, the faster neutrinos can travel, the more of the Universe we can see with neutrinos.  Thus the flux intensity of the diffuse background of all neutrino sources in the Universe should increase linearly with the neutrino speed $v$, in accordance with Olber's Paradox.  However, there already are limits on the diffuse neutrino background, and some of these are already near theoretical predictions for the neutrino background assuming $v = c$.  The fact that we do not see a neutrino background at high energies therefore implies that $v$ is not many orders of magnitude higher than $c$ at these energies.  

\section{Explanation and Assumptions}
The idea behind this limit is that superluminal neutrinos increase the intensity (defined here as number flux per steradian) of the diffuse background.  Suppose neutrino sources have equal luminosity and uniform number density throughout the Universe.  In a shell of thickness $dR$ at distance $R$ from Earth, the flux of neutrinos per source goes as $R^{-2}$ but the number of sources in the shell goes as $R^2 dR$.  Thus each shell has the same flux as viewed from Earth, and the integrated flux increases linearly with the maximum observable distance.  Neutrinos with $v \gg c$ would be observable out to a larger horizon, and the neutrino background would have a correspondingly higher intensity.

Note the number density of the neutrino background is \emph{not} increased: that is set simply by the rate of neutrino injection per comoving volume (ultimately from the energetics of neutrino production) and the age of the Universe.  Rather, the higher speed of neutrinos increases the rate at which the neutrinos hit a detector on the Earth.  Thus, models in which the neutrinos transform into some intermediate state which is superluminal and then turn back into subluminal particles before hitting the detector (e.g., \cite{DarkFTLNu}) are not constrained --- those models alter which sources contribute to the background but not its intensity.  

The number density spectrum $dn/dE$ [$\cm^{-3} \GeV^{-1}$] of neutrinos is calculated as $\int dt \rho_{\rm com} [dQ/dE^{\prime}] [dE^{\prime}/dE]$, where $dt = dz / [H_0 (1 + z) \sqrt{\Omega_{\Lambda} + \Omega_M (1 + z)^3}]$ describes the lookback time, $\rho_{\rm com}$ is the comoving density of sources, and $dQ/dE^{\prime}$ is the source-frame number luminosity [$s^{-1} \GeV^{-1}$] of a source.  I have assumed that neutrinos accumulate over cosmic time in the FRW metric, rather than some other coordinate time.  From the number density spectrum, we can calculate the present-day background intensity simply as $d\Phi/dE = dn/dE \times v / (4\pi)$, where $v$ is the current neutrino speed at energy $E$, to get
\begin{equation}
\frac{d\Phi}{dE} = \frac{v}{4 \pi H_0} \int \frac{dz}{(1 + z) \sqrt{\Omega_{\Lambda} + \Omega_M (1 + z)^3}} \rho_{\rm com} \frac{dQ}{dE^{\prime}} \frac{dE^{\prime}}{dE}.
\end{equation}

I will assume that $dE^{\prime}/dE = (1 + z)$; this holds for particles with $E \approx pc (1 + \delta)$, which leads to $dE/dp = v \approx c (1 + \delta)$.  The redshift effects are relatively minor unless the energy dependence is much more extreme than $E^{\prime} \propto (1 + z)$ \cite{OlbersSolution}.  It can then be shown that $d\Phi/dE$ scales directly with the constant superluminal velocity at all energies: thus as $v \to \infty$, $d\Phi/dE \to \infty$ in accordance with Olber's Paradox.

However, some caveats apply to these limits:
\begin{enumerate}
\item As discussed above, I assume that neutrinos themselves are superluminal in the space the detector occupies, because this is a limit on flux and not density.
\item I assume that expected astrophysical neutrino backgrounds from pion production processes actually exist; if pions do not decay or neutrinos decay en route \citep{KinematicLimits,NuToePair,NuVsPionDecay,DecayOfAstroNus}, the limits will not apply.
\item I assume that the neutrino speed is constant at every point in the Universe with respect to the frame in which the cosmic expansion is isotropic.  If this is not the case, there is a preferred location in the Universe.  Elsewhere, the diffuse neutrino background will appear anisotropic and superluminal neutrinos may come from our future with respect to cosmic time (from $-1 < z < 0$).  Furthermore, the neutrino density would represent the integral of energy injection over some coordinate time that may bear little resemblance to cosmic time, and the neutrino density could then be far lower than expected.  
\item Likewise, I ignore the effects of the Earth's motion $v_{\oplus}$ to this frame, which will introduce large anisotropy in the Earth-frame speed if $v / c \ga c / v_{\oplus} \approx 800$ for $v_{\oplus} = 370\ \kms$ \cite{CMBDipole}.  The Earth-frame number density will also appear to be anisotropic, with $dn/d\theta \to 0$ as $v(\theta) / c \to \infty$ (because neutrinos arrive the moment they are emitted in Earth-frame and there are only a finite number of neutrinos emitted within the neutrino horizon) so the measured event rate is not changed to first order.  
\item I assume that the region of the Universe with neutrino sources is infinitely large: if it is not, the enhancement of the background will reflect the true size of the neutrino-emitting Universe and not the neutrino horizon (although this may be interesting in its own right).  
\item Finally, I assume the Universe is infinitely transparent to neutrinos.  However, at energies $\ga 10^{22}\ \eV$, neutrinos may be absorbed over distances much greater than $c / H_0$ through interactions with low energy neutrinos and photons \cite{NuOpacity}.\footnote{A similar absorption-limited background was proposed by Olber as the explanation for why the night sky is dark in optical light.}  
\end{enumerate}

\section{Application}
While we have not detected TeV or higher energy neutrinos from astrophysical sources, we expect that neutrino sources exist because of the observations of cosmic ray (CR) protons and nuclei.  These particles create pions through either collisions with ambient nucleons ($pp$) or photons ($p\gamma$).  Then, assuming the kinematics are not altered, the pions decay into secondary electrons and positrons, $\gamma$-rays, and neutrinos.  Thus wherever there are CRs and enough gas or radiation (of sufficient energy for the $p\gamma$ process to occur), we expect neutrinos to be emitted.  

I consider GeV--PeV neutrinos from star-forming galaxies and PeV--ZeV neutrinos from extragalactic UHECRs, which both probably exist, though the exact levels of these backgrounds is still unclear.  In addition, there may be intense neutrino backgrounds from active galactic nuclei \cite{AGNNu} and other sources, but the existence of these backgrounds is more uncertain.  Finally, I note that the MeV diffuse supernova neutrino background is expected to have an intensity only a factor of $\sim 4$ below current limits.  Thus $v/c \la 4$ at $\sim 10 - 30\ \MeV$ \cite{DSNB}, although these limits are unfavorable compared to extant ones from SN1987A \cite{SN1987ALIVLimit} and the OPERA experiment at GeV energies \cite{Adam11}.

\subsection{The Star-Forming Galaxy Neutrino Background}
The Milky Way glows in pionic $\gamma$-ray emission, indicating the presence of CR protons interacting with gas throughout the Galactic disk \cite{GalacticGamma}.  The detection of synchrotron radio emission \cite{FRC} and now GeV $\gamma$-ray emission in other nearby galaxies \cite{NormGalGamma,StarburstGamma} indicates that CRs are also accelerated in them as well, and furthermore, that the CR energy injection rate is related to the star-formation rate.  This led to the prediction of a minimum ``guaranteed $\gamma$-ray background'' from $pp$ collisions throughout galaxies in the Universe \cite{SFGGammaBack,Thompson07}, but the same process also produces a neutrino background \cite{Stecker79,Loeb06}.  At high energies the neutrino spectrum of a galaxy is proportional to, and of roughly the same intensity as its $\gamma$-ray spectrum \cite{Stecker79}.  I consider two subclasses of star-forming galaxies: ``normal'' (Milky Way-like) galaxies and starburst galaxies.

In normal galaxies, CR protons escape by diffusion, leading to rapidly falling steady-state CR proton, $\gamma$-ray, and neutrino spectra ($dN/dE \propto E^{-2.7}$; \cite{Ginzburg76}).  The spectrum steepens at a few PeV (the ``knee''); since the typical energy of a neutrino is about 5\% of the proton energy, this translates to a cutoff in the neutrino spectrum at $\sim 100\ \TeV$ \cite{Loeb06}.   I therefore assume a $E^{-2.7}$ power law spectrum for the Milky Way, with a cutoff at 100 TeV.  Ref. \cite{Strong10} modeled CR propagation in the Milky Way and found a $\ge 1\ \GeV$ pionic $\gamma$-ray luminosity of $3.7 \times 10^{38}\ \ergps$.  Scaling to the star-formation rate of the Milky Way ${\rm SFR}_{\rm MW} = 2\ \Msun\ \yr^{-1}$ (e.g.,\cite{MWSFR}), I use a neutrino spectrum of 
\begin{equation}
\frac{dQ}{dE^{\prime}} = 2.1 \times 10^{35} \sec^{-1} \GeV^{-1} \left(\frac{E^{\prime}}{\GeV}\right)^{-2.7} \exp\left(\frac{-E^{\prime}}{100\ \TeV}\right) \left(\frac{\rm SFR}{\Msun\ \yr^{-1}}\right).
\end{equation}

Starburst galaxies have recently been detected in GeV--TeV gamma-rays \cite{StarburstGamma}, and these gamma-rays are believed to be pionic \cite{StarburstModels,M82N253}.  The high star-formation rate in the starburst regions leads to high cosmic ray energy densities; when combined with the high gas densities in the starbursts, high pionic luminosities are expected \cite{BrightSBPion,Thompson07}.  I assume a $E^{-2.2}$ spectrum for starburst galaxies as observed, and scale to the $\ge 1\ \GeV$ luminosity of M82 ($1.9 \times 10^{40}\ \ergps$; \cite{M82N253}) and M82's star-formation rate ${\rm SFR}_{\rm M82} = 8\ \Msun\ \yr^{-1}$ from its infrared luminosity \cite{M82IRtoSFR} \footnote{As converted into a ``Salpeter A'' initial mass function of stars, as used in \cite{Hopkins06}}: 
\begin{equation}
\frac{dQ}{dE^{\prime}} = 7.6 \times 10^{35} \sec^{-1} \GeV^{-1} \left(\frac{E^{\prime}}{\GeV}\right)^{-2.2} \exp\left(\frac{-E^{\prime}}{100\ \TeV}\right) \left(\frac{\rm SFR}{\Msun\ \yr^{-1}}\right).
\end{equation}

The cosmic star-formation rate is given in ref. \cite{Hopkins06}.  In order to calculate the spectrum of the neutrino background, the fraction of star-formation in normal galaxies versus starbursts is necessary.  While debated in the literature \cite{SBFracDebate}, I use a conservative 5\% at all redshifts in starburst galaxies \cite{LowSBFrac}, with the rest in the fainter normal galaxies.  Note that I ignore the effects of gas evolution, which can enhance the neutrino flux in normal galaxies \cite{Loeb06,Fields10}.  I assume $\Omega_M = 0.3$, $\Omega_{\Lambda} = 0.7$, and $H_0 = 70\ \kms\ \Mpc^{-1}$ for this calculation.

\begin{figure}
\centerline{\includegraphics[width=15cm]{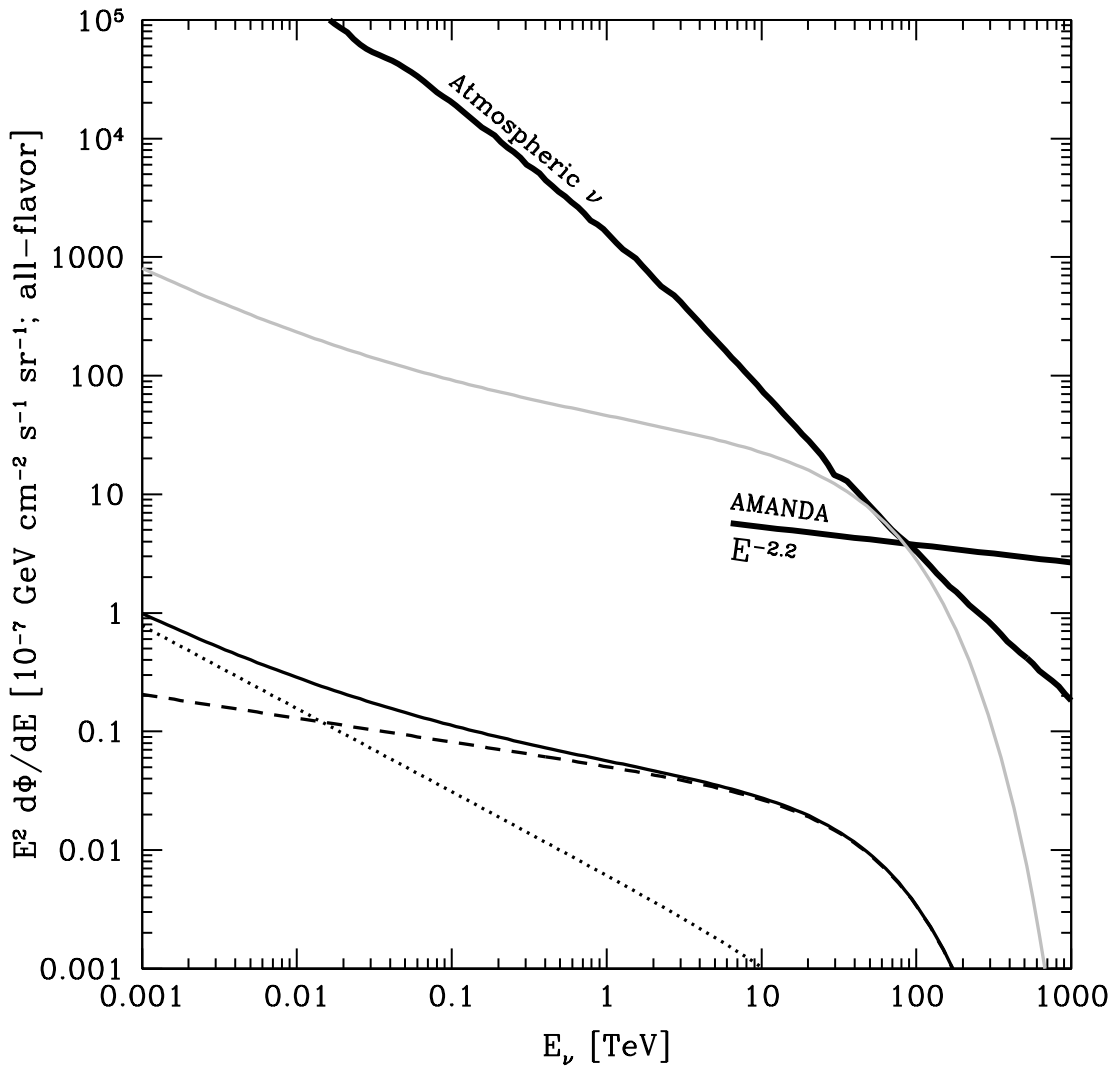}}
\caption{Predicted pionic neutrino fluxes from star-forming galaxies with $v/c = 1$ (black) compared to the atmospheric neutrino spectrum (thick black; \cite{GaisserHonda02}) and the AMANDA limits on a $E^{-2.2}$ spectrum from 6.3 TeV to 1.3 PeV (thick black; \cite{AMANDAStarbursts}).  The dotted line is the normal galaxy contribution and the dashed line is the starburst contribution (if 5\% of the cosmic star formation rate is in starbursts).  The grey line is the background scaled up by the maximum allowed $v/c = 820$. \label{fig:FTLNuSFGBounds}}
\end{figure}

Current neutrino limits with IceCube are for an $E^{-2}$ spectrum and only apply at energies of $35\ \TeV - 7\ \PeV$ \cite{NewIceCubeLimit}, where the predicted neutrino spectrum tails off (Figure~\ref{fig:FTLNuSFGBounds}).  However, the star-formation neutrino background is certainly smaller than the atmospheric neutrino background, which is measured up to 400 TeV \cite{AtmosNuObs}.  I therefore take the ratio of the atmospheric neutrino spectrum from \cite{GaisserHonda02} and the predicted neutrino background to find a constraint on $v / c$.  This is very conservative, since pionic astrophysical neutrino backgrounds have different spectra, flavor ratios, and angular distributions than atmospheric neutrinos.

\begin{table*}
\caption{Summary of constraints on $v/c$ of neutrinos of energies from 30 GeV to 300 TeV from star-formation neutrino background. At 100 TeV and above, the neutrino background depends on where the ``knee'' in the CR nuclei spectrum is.  I conservatively compare the spectra to the atmospheric neutrino background, and also compare the starburst component alone to the $E^{-2.2}$ limit from AMANDA.\label{table:SFGNu}}
\begin{tabular}{lcccc}
        & \multicolumn{3}{c}{Limit on $v/c$ (Atmospheric)} & Limit on $v/c$ ($E^{-2.2}$ AMANDA)\\
Energy  & Normal Galaxies & Starbursts & Total & Starbursts\\
\hline
30 GeV  & $7.6 \times 10^5$ & $5.3 \times 10^5$ & $3.1 \times 10^5$ & \nodata\\
100 GeV & $6.5 \times 10^5$ & $2.5 \times 10^5$ & $1.8 \times 10^5$ & \nodata\\
300 GeV & $4.3 \times 10^5$ & $9.5 \times 10^4$ & $7.8 \times 10^4$ & \nodata\\
1 TeV   & $2.7 \times 10^5$ & $3.2 \times 10^4$ & $2.8 \times 10^4$ & \nodata\\
3 TeV   & $1.5 \times 10^5$ & $1.1 \times 10^4$ & $1.0 \times 10^4$ & \nodata\\
10 TeV  & $7.4 \times 10^4$ & $2900$            & $2800$            & $100$\\
30 TeV  & $4.4 \times 10^4$ & $990$             & $970$             & $150$\\
100 TeV & $7.4 \times 10^4$ & $980$             & $960$             & $550$\\
300 TeV & $6.5 \times 10^5$ & $5500$            & $5400$            & $1.1 \times 10^4$\\
\end{tabular}
\end{table*}

These constraints on $v / c$ are weak, especially at low energies where the atmospheric background is high (Figure~\ref{fig:FTLNuSFGBounds} and Table~\ref{table:SFGNu}).  At 100 GeV, $v / c \la 2 \times 10^5$, but the best constraint is at 63 TeV, where $v / c \la 820$.  The constraints would be stronger if the starburst fraction were higher (the starburst contribution already dominates above 12 GeV); if it were 100\%, the best constraint would be $v/c \la 42$ at 63 TeV (compared to $\sim 340$ from a naive quadratic extrapolation of the OPERA result).  

However, as IceCube sets increasingly stringent limits on $E^{-2.2}$ neutrino backgrounds at 10 TeV, the limits on $v/c$ should decrease as the starburst neutrino background approaches detection.  AMANDA set a limit on $E^{-2.2}$ spectra extending from 6.3 TeV to 1.3 PeV \cite{AMANDAStarbursts}.  Comparing to my predicted starburst spectrum and ignoring the fact that its energy cutoff would alter the neutrino analysis, I find $v / c \la 90$ at 6.3 TeV.  

\subsection{The UHECR Neutrino Background}
Ultrahigh energy cosmic rays (UHECR) are observed at energies up to $10^{20}\ \eV$ and beyond, and are generally believed to be extragalactic in origin, although their composition is unknown.  As they traverse the intergalactic medium, UHECR nucleons interact with the CMB to produce pions above the GZK (Greisen-Zatsepin-Kuzmin) threshold ($\sim 10^{20}\ \eV$; \cite{GZK}); these presumably ultimately decay into neutrinos and other particles \cite{GZKNuPred}.  The existence of the photopion process is supported by the suppression of the UHECR flux above 40 EeV \cite{GZKObs}, although photodisintegration of heavy ions could also create this dip \cite{PhotoDisint}.  

A number of experiments have sought these GZK neutrinos, and stringent limits have been placed, ruling out some models already (e.g., \cite{AMANDA-UHE,IceCube-UHE,Auger-UHE,RICE-UHE,ANITA-UHE}).  Even in the most pessimistic models where UHECRs consist of entirely heavy nuclei, some neutrinos are expected from photopion production on the extragalactic background light in the infrared, as well as photopion production from secondary protons after photodisintegration of heavy nuclei \citep{Kotera10}.  A large $v/c$ at EeV energies would enhance the flux from these small backgrounds to the point where they are detectable.   

\begin{figure}
\centerline{\includegraphics[width=15cm]{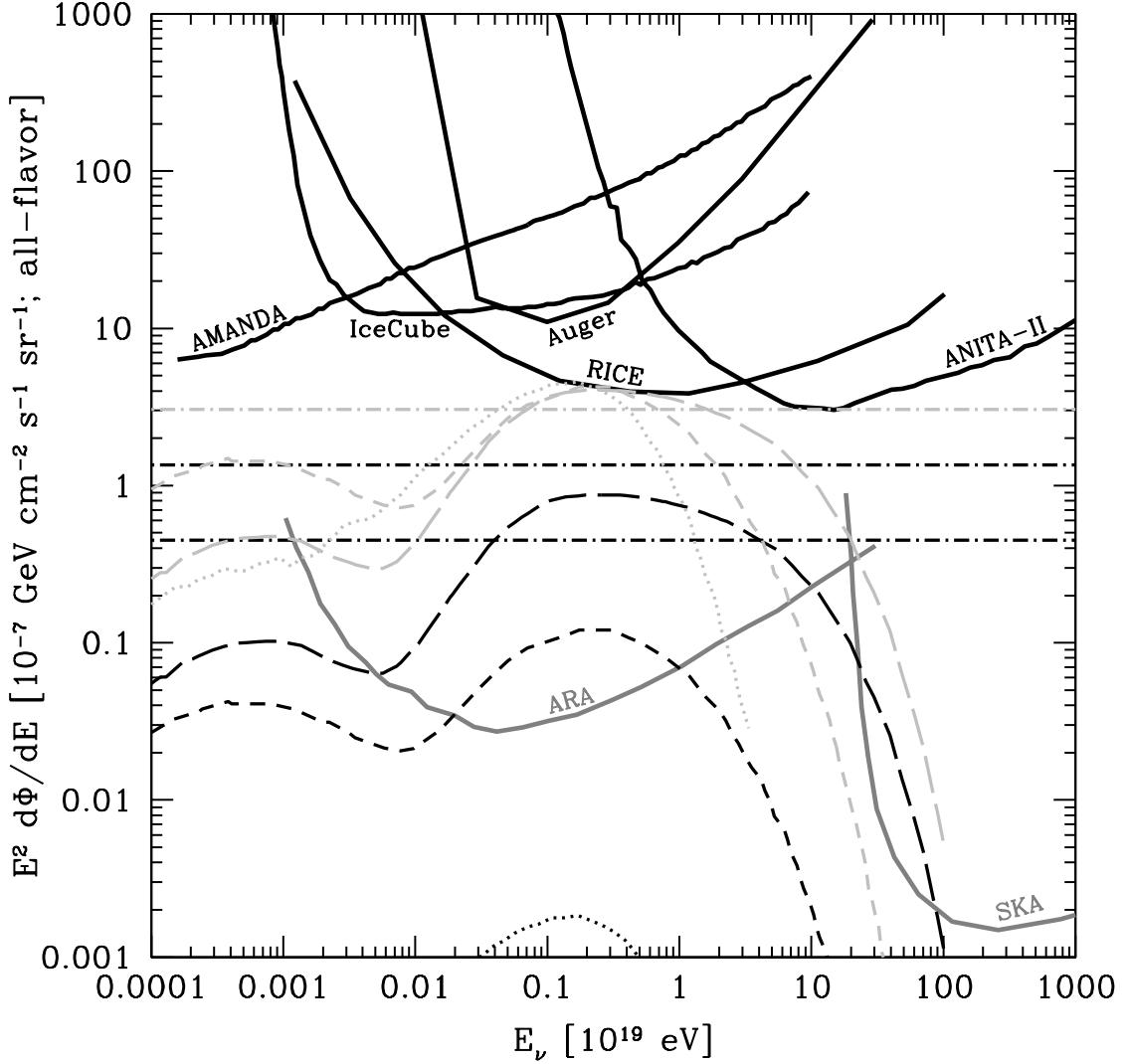}}
\caption{Predicted neutrino fluxes from UHECRs with $v/c = 1$ (black) compared to instrument sensitivities (thick solid black). The dotted line is a pessimistic pure iron composition model, the short-dashed line is a typical proton dip model, and the long-dashed line is a proton model with strong evolution from Ref. \citep{Kotera10}.  The dash-dotted lines are the Waxman-Bahcall bounds on the neutrino background \cite{WBLimit}.  In grey, I show the models scaled up by the maximum $v/c$ allowed by the data.  Observed limits are from AMANDA \cite{AMANDA-UHE,IceCube-UHE}, IceCube \cite{IceCube-UHE}, Auger \cite{Auger-UHE}, RICE \cite{RICE-UHE}, and ANITA \cite{ANITA-UHE}. In grey, I also show the projected limits from ARA \cite{ARA} and SKA \cite{NuMoonSKA}.\label{fig:FTLNuGZKBounds}}
\end{figure}

I consider three models of the UHE neutrino flux from \cite{Kotera10}, who consider a wide range of parameters.  In their most optimistic case, UHECRs are protons and injection evolves strongly with redshift (as radio galaxies do), in a more typical case, UHECRs are protons and evolve more weakly with redshift (as star formation does), and in a pessimistic case, UHECRs are iron nuclei and injection does not evolve with redshift (Figure~\ref{fig:FTLNuGZKBounds}).  

Using these predictions, I find that the best constraint of all energies is $v/c \la 2500$ at 0.18 EeV in the pessimistic model, $v/c \la 35$ at 3.2 EeV in the typical model, and $v/c \la 4.6$ at 4.2 EeV in the optimistic model.  These constraints are particularly interesting in that a naive quadratic extrapolation of the OPERA result would indicate $v/c \approx 9\times 10^{10} (E / 1\ {\rm EeV})^2$; even a linear extrapolation gives $v/c \approx 1500 (E / 1\ {\rm EeV})$.  

\begin{table*}
\caption{Summary of constraints on $v/c$ of neutrinos of energies from 100 GeV to 100 TeV from UHECR neutrino background.  The GZK models are from ref. \cite{Kotera10}: the optimistic model has a proton composition and strong evolution; the medium case has a proton composition and weaker evolution; the pessimistic case has an iron composition. Also included are the bounds if the neutrino injection rate is at the Waxman-Bahcall limit \cite{WBLimit}.\label{table:UHENu}}
\begin{tabular}{lccccc}
        & \multicolumn{5}{c}{Limit on $v/c$}\\
        & \multicolumn{3}{c}{GZK Models}    & \multicolumn{2}{c}{Waxman-Bahcall}\\
Energy  & Optimistic & Medium & Pessimistic & Non-Evolving & With Evolution\\
\hline
3 PeV   & 75      & 170               & $6.2 \times 10^4$ & 15  & 5.0\\
10 PeV  & 100     & 270               & $7.8 \times 10^4$ & 23  & 7.8\\
30 PeV  & 220     & 570               & $6.9 \times 10^4$ & 35  & 12\\
100 PeV & 130     & 570               & $2.7 \times 10^4$ & 27  & 9.1\\
300 PeV & 28      & 190               & 9700              & 21  & 7.1\\
1 EeV   & 6.6     & 52                & 3100              & 12  & 3.9\\
3 EeV   & 4.9     & 35                & 2800              & 9.4 & 3.1\\
10 EeV  & 5.2     & 56                & $1.1 \times 10^4$ & 8.6 & 2.9\\
30 EeV  & 8.7     & 220               & $2.9 \times 10^5$ & 10  & 3.4\\
100 EeV & 13      & 1500              & \nodata           & 7.0 & 2.3\\
300 EeV & 82      & $6.5 \times 10^4$ & \nodata           & 8.2 & 2.7\\
1 ZeV   & \nodata & \nodata           & \nodata           & 11  & 3.6\\
10 ZeV  & \nodata & \nodata           & \nodata           & 25  & 8.3\\
100 ZeV & \nodata & \nodata           & \nodata           & 88  & 29\\
\end{tabular}
\end{table*}

Finally, a characteristic density for the UHE neutrino background is the Waxman-Bahcall bound, in which the energy injection rate in UHE neutrinos is less than that of observed UHECRs \cite{WBLimit}.  If the UHE neutrino energy injection rate is at the Waxman-Bahcall limit, then $v/c$ is less than 90 (if there is no redshift evolution) or 30 (with redshift evolution) over the entire range from 2 PeV to 100 ZeV.  The most stringent limit would then be at 140 EeV: $v/c < 6.8$ without evolution or $v/c < 2.3$ with evolution.  However, the neutrino background density could easily be much lower than the Waxman-Bahcall limit, which would weaken these limits.  

\section{Conclusion}
I have argued that if neutrino are superluminal and stable, it is conceivable that $v \gg c$ at TeV energies and beyond.  In standard Lorentz invariance violation models, this possibility is ruled out \cite{NuToePair,NuToePairPRL,ICARUS,KinematicLimits,NuSplitting}, but some have suggested that there could be ways around these limits \cite{DarkFTLNu,FTLinDSR,Bimetric}.  Astrophysical neutrinos could provide strong tests of superluminal neutrino speeds, especially through the timing of neutrinos arriving from transients, but no astrophysical neutrinos of these energies have been detected yet.  

By taking advantage of the increased neutrino horizon size if $v \gg c$, I constrain $v / c$ on TeV--ZeV scales.  The lack of neutrinos from star-forming galaxies conservatively constrains $v/c \la 820$ at 60 TeV and the lack of GZK neutrinos conservatively limits $v/c \la 2500$ at 0.1 EeV energies.  The limits are subject to several caveats: most notably, pionic neutrinos must actually be produced and survive, and must actually be superluminal when they cross the detector since the limit is on background intensity and not density.   

The limits can be improved by further understanding of neutrino sources, so that we have a more accurate baseline prediction to compare the limits to.  A determination of the composition of UHECRs \cite{UHECRComposition} and constraints from the cascade radiation of GZK photons (e.g., \cite{GZKPhotons}) will allow us to better predict what the ``guaranteed'' GZK neutrino flux is.  Likewise better understanding of the neutrino background from star-forming galaxies may come from studies of the accompanying pionic gamma-ray background. 

On the observational side, experiments such as IceCube, the Askaryan Radio Array (ARA; \cite{ARA}), and the Square Kilometer Array (SKA) will push down the limits on the neutrino background, with ARA expected to detect the UHE neutrino background except in the most pessimistic cases.  Future neutrino detections of sources like AGNs may demonstrate additional ``guaranteed'' neutrino sources.  Eventually, if the uncertainties in the predicted neutrino background are less than a factor of 2 and the neutrino background is detected at predicted levels, we will be able to probe directly the regime where $\Delta v \approx c$ where any change in energy dependence might naturally occur.  

Of course, the general idea behind this method is not just limited to neutrinos, but could apply to any particle.  On the other hand, just as high order effects might cause $v$ to increase rapidly at high energies, by the same logic, high energy neutrinos may travel far slower than $c$.  Extreme subluminal speeds at high energies would have opposite effects as superluminal speeds: the background flux (but not density) would be much lower than predicted.  A smaller horizon is already predicted for the relic cosmic neutrino background, where the neutrino mass is important \cite{CNBHorizon}.  

If it turns out that $v$ really is much larger than $c$, a number of interesting consequences arise in neutrino astronomy.  We would see photon-observed galaxies as they are now in cosmic time, millions or even billions of years after their light was emitted, in completely different stages of their evolution.  Since there is less star formation \cite{Hopkins06} and active galactic nucleus activity \cite{AGNEvol} at $z \approx 0$, we would see far fewer active galaxy neutrino sources and gamma ray burst neutrino flares within the photon horizon, even in a blind survey not affected by long delays.  The diffuse neutrino background is brighter only because of the contributions of neutrinos emitted at $z \ga 1$ from sources far beyond the photon horizon that individually have much lower observed fluxes.  In contrast, if $v \ll c$, then we would see sources near the neutrino horizon as they were at high redshift, when they presumably were brighter, even though the background as a whole is smaller.

\acknowledgments
I would like to acknowledge discussions with John Beacom and Todd Thompson.  BCL is supported by a Jansky fellowship from the NRAO.  NRAO is operated by Associated Universities, Inc., under cooperative agreement with the National Science Foundation.

\end{document}